\documentclass[noshowpacs,amsmath,twocolumn,superscriptaddress,8pt,aps,prb]{revtex4-1} % this article uses the Revtex 4-1 format
\usepackage{setspace}
\usepackage{amsmath}
\usepackage{breqn}
\usepackage{graphicx}
\usepackage[nearskip,margin = 0pt]{subfig}
\usepackage{verbatim} % for commenting
\usepackage{amsfonts}
\usepackage{amssymb}
\usepackage{epstopdf}
\usepackage{xcolor}
\DeclareGraphicsExtensions{.pdf,.eps,.png,.jpg,.mps}

\begin{document}
\title{Bridging ultra-high-Q devices and photonic circuits}
\author{Ki Youl Yang$^1$*, Dong Yoon Oh$^1$*, Seung Hoon Lee$^1$*, Qi-Fan Yang$^1$, Xu Yi$^1$ and Kerry Vahala$^{1\dagger}$\\
$^1$T. J. Watson Laboratory of Applied Physics, California Institute of Technology, Pasadena, California 91125, USA.\\
*These authors contributed equally to this work.\\
$^{\dagger}$Corresponding author: vahala@caltech.edu}
 
 \maketitle

{\bf Optical microcavities are essential in numerous technologies and scientific disciplines. However, their application in many areas relies exclusively upon discrete microcavities in order to satisfy challenging combinations of ultra-low-loss performance (high cavity-Q-factor) and cavity design requirements. Indeed, finding a microfabrication bridge connecting ultra-high-Q device functions with micro and nanophotonic circuits has been a long-term priority of the microcavity field. Here, an integrated ridge resonator having a record Q factor over 200 million is presented. Its ultra-low-loss and flexible cavity design brings performance that has been the exclusive domain of discrete silica and crytalline microcavity devices to integrated systems. Two distinctly different devices are demonstrated: soliton sources with electronic repetition rates and high-coherence Brillouin lasers. This multi-device capability and performance from a single integrated cavity platform represents a critical advance for future nanophotonic circuits and systems.}

Optical microcavities\cite{Vahala2003} provide diverse device functions that include frequency microcombs\cite{Pascal2007,Kippenberg2011}, soliton mode-locked microcombs\cite{Herr2014,Xu2015,Brasch2014,Joshi:OL:2016,Wang:OE:2016}, Brillouin lasers\cite{Tomes2009,grudinin2009brillouin,Hansuek2012,Eggleton:2017,Rakich2017Laser},  bio and nano-particle sensors\cite{vollmer2008,Lu2011,Vollmer2012}, cavity optomechanical oscillators\cite{Kippenberg2008}, parametric oscillators\cite{Kippenberg2004,Savchenkov2004}, Raman lasers\cite{Spillane2002}, reference cavities/sources\cite{Matsko:07,Alnis2011,Lee:NatCommun:2013,Loh2015} , and quantum optical devices\cite{Aoki2006,Kimble2008}. Key performance metrics improve with increasing Q-factor across all applications areas\cite{Vahala2003}. For example, higher Q factors dramatically reduce power consumption as well as phase and intensity noise in signal sources, because these quantities scale inverse quadratically with Q factor. Also, higher Q improves the ability to resolve a resonance for sensing or for frequency stabilization. Such favorable scalings of performance with Q factor have accounted for a sustained period of progress in boosting Q factor by reducing optical loss in resonators across a range of materials \cite{Xiong:NanoLetters:2012,Moss_2013,Hausmann:NaturePhotonics:2014, Lu:SciRep:2015,Lu:Optica:2016}. Likewise, the need for complex microcavity systems that leverage high-Q factors has driven interest in low-loss monolithically integrated resonators\cite{Lipson_2010,Ramiro2012,Moss_2013,Jung:OL:2013,Bowers_2014,Hausmann:NaturePhotonics:2014, Xuan2016,Pfeiffer2016,Ji2017}. For example, Q values in waveguide-integrated devices to values as high as 80 million\cite
{Bowers_2014} and 67 million\cite{Ji2017} in strongly-confined resonators have been attained. 

Nonetheless, the highest Q-factor resonators remain discrete devices that are crystalline\cite{Grudinin2006} or silica based\cite{Vahala2003,Armani2003,Papp2011,Hansuek2012}. These discrete resonators are moreover unique in the microcavity world in terms of overall performance and breadth of capability. This includes generation of electronic-repetition-rate soliton streams as required in optical clocks \cite{papp2014,soltani2016enabling,Frank2017} and optical synthesizers\cite{Spencer2017}, rotation measurement at near-earth-rate sensitivity in micro-optical-gyros\cite{Li:Optica:17,Liang:Optica:2017}, synthesis of high-performance microwave signals\cite{maleki2011,Li2013,Li2014,Liang2015}, and operation as high-stability optical frequency references\cite{Matsko:07,Alnis2011,Lee:NatCommun:2013} and reference sources\cite{Loh2015}. Functions such as these belong to a new class of compact photonic systems that rely upon ultra-high-Q fabrication methods which have so far defied photonic integration. 

In this work a monolithic microcavity having a record high Q factor is demonstrated. The materials, process steps and, in particular, the use of a PECVD silicon nitride waveguide enable full integration of these ultra-high-Q resonators with other photonic devices. Critically, and as required for new system-on-a-chip applications, the resonator supports design controls required to realize many device functions previously possible using only discrete (non-waveguide-integrated) devices. To demontrate its capability, electronics-rate solitons (15 GHz) are generated at a low pumping power level. Also, as a distinctly different capability, high-coherence stimulated Brillouin laser oscillation is demonstrated. Beyond the necessity of ultra-high-Q factor in these demonstrations, each device presents additional challenges. The soliton source requires resonator dispersion control to support dissipative Kerr solitons and the Brillouin laser requires precise diameter control to phase match the Brillouin laser process. As representative photonic circuits based on these devices, Fig. 1a conceptualizes a dual Brillouin optical gyroscope and a dual-comb spectrometer modeled after recent demonstrations using discrete ultra-high-Q devices \cite{Li:Optica:17,Suh2016}. 

%These systems represent a large opportunity for miniaturization if Connecting these systems with a , but they present serious challenges for combination with ultra-high-Q through a scalable nanofabrication fabrication process.  
 
\begin{figure*}[t!]
\centering
\includegraphics[width=1.0\linewidth]{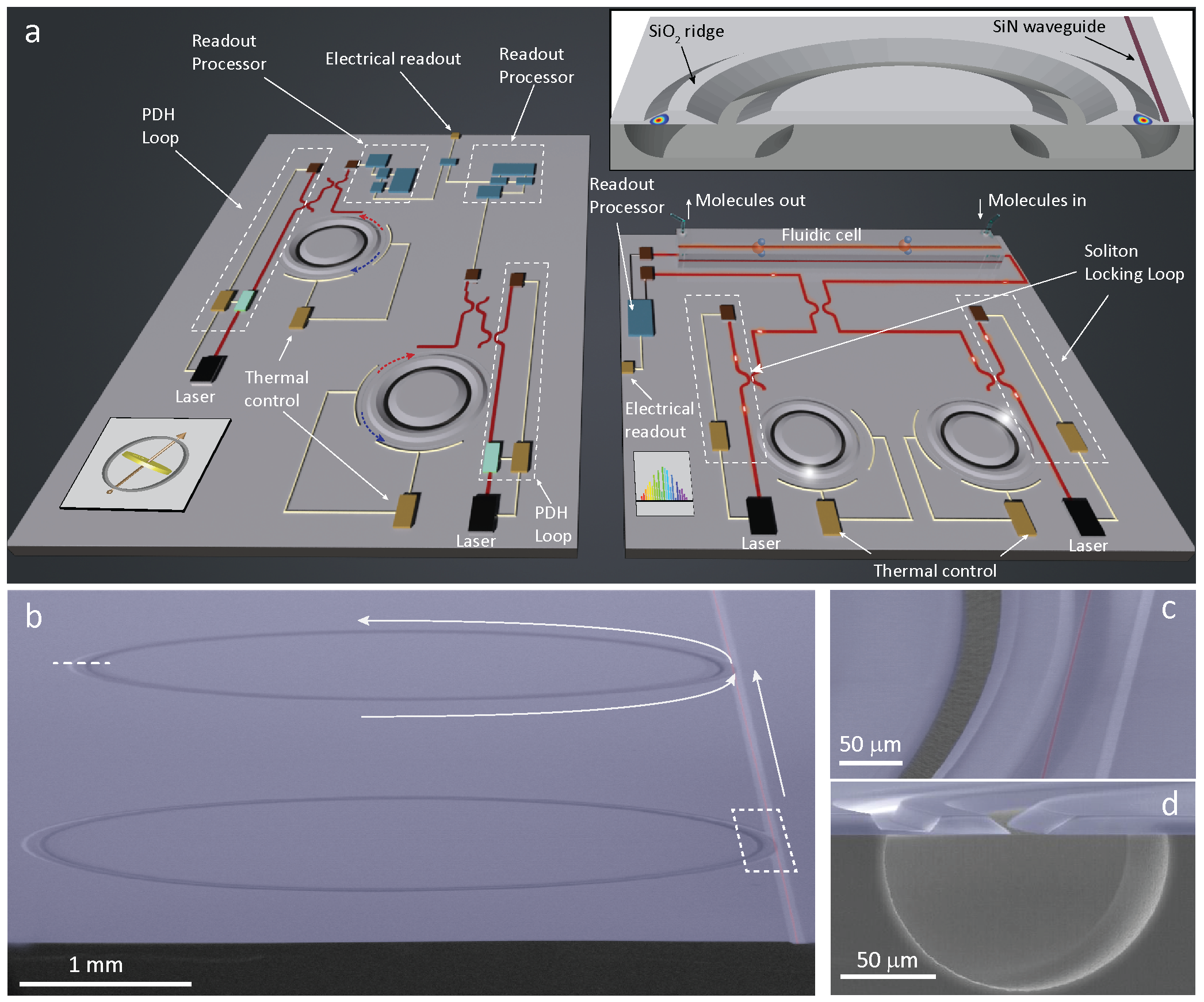}
\captionsetup{singlelinecheck=no, justification = RaggedRight}
\caption{{\bf{Integrated ultra-high-Q microresonator}} (\textbf{a}) Artists conception of a dual-gyro chip (left) and micro-fluidic \cite{Quake2005} dual-comb spectrometer (right) based on recent discrete Brillouin gyro\cite{Li:Optica:17} and dual-comb spectroscopy demonstrations\cite{Suh2016}. Component color codings are as follows: silicon nitride waveguides (red lines), electrical traces (gold lines), laser sources (black), detectors (brown), modulators (turquoise), loop-control electronics (gold), and readout processing electronics (blue). Both chips include Pound-Drever-Hall (PDH) locking loops and thermal control for the resonators. Note: gyros are counter-pumped to reduce possible systematic errors. First and second Brillouin Stokes waves (blue and red) are shown in the gyroscope resonators. Soliton pulses (white) appear in the spectrometer resonators.  Inset: Rendering showing silica ridge resonator with nearby silicon nitride waveguide. Fundamental spatial mode is illustrated in color. (\textbf{b}) SEM image of two ridge resonators with a common silicon nitride waveguide (false red color). White arrows show the direction of circulation within the top resonator for the indicated direction of coupling from the waveguide. Dashed white box is the region for the zoom-in image in 1c. Dashed white line segment gives the location of cleavage plane used for preparation of the SEM image in 1d.  (\textbf{c}) SEM zoom-in image of the waveguide-resonator coupling region shown within the dashed white box in 1b.  (\textbf{d}) SEM image of resonator cross section prepared by cleaving at the dashed white line in 1b.}
\label{fig:Fig1}
\end{figure*}

\begin{figure*}[t!]
\centering
\includegraphics[width=1.0\linewidth]{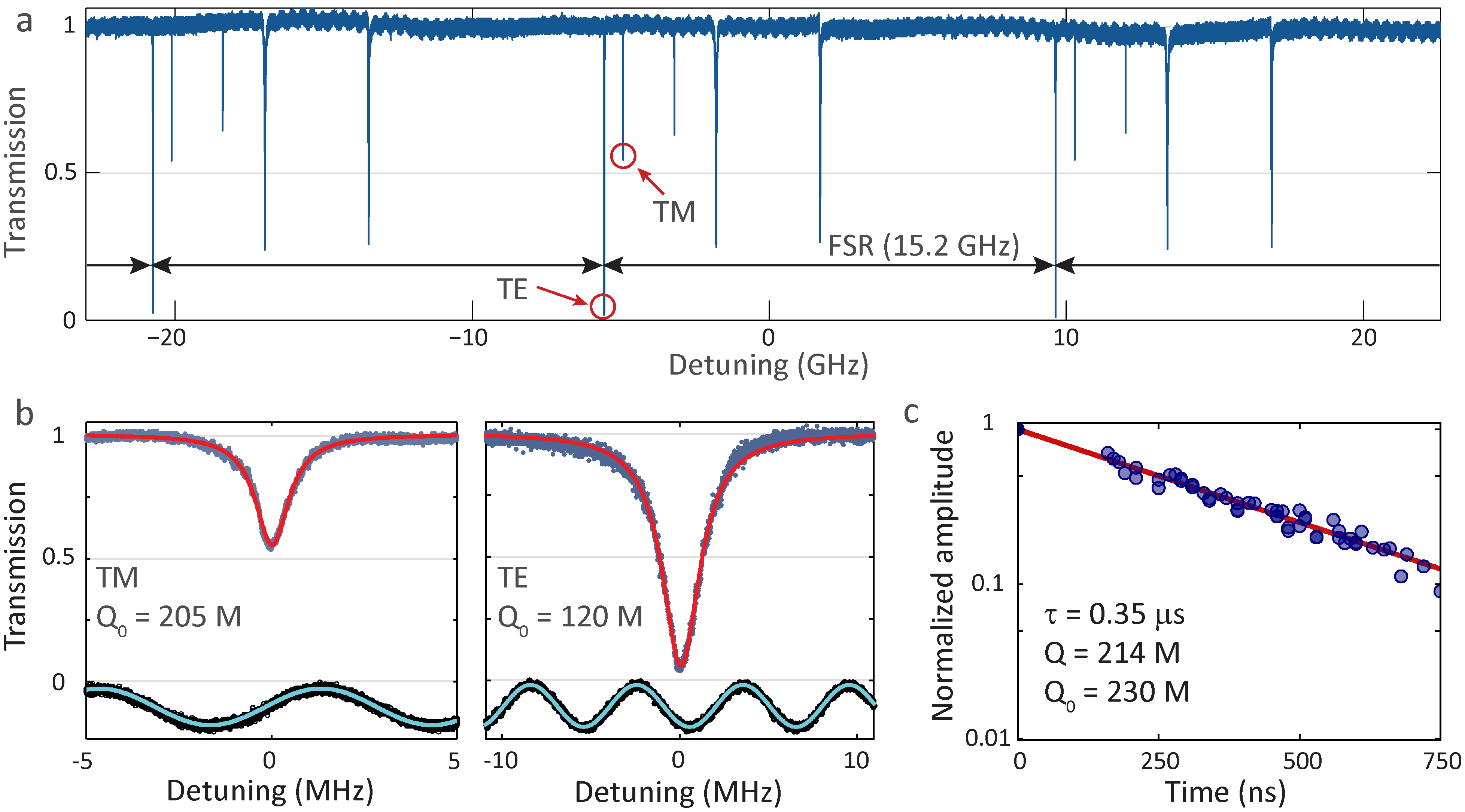}
\captionsetup{singlelinecheck=no, justification = RaggedRight}
\caption{{\bf{Spectral scan of integrated ridge resonator and ring-down measurement}} (\textbf{a})  Spectral scan encompassing three free-spectral-ranges. Fundamental TE and TM modes are indicated. (\textbf{b}) High resolution zoom-in scan for the fundamental TM and TE modes with intrinsic quality factor ($Q_o$) indicated (M = million). Red curve is a Lorentzian fit. The green sinusoidal signal is a frequency calibration scan using a radio-frequency calibrated fiber Mach-Zehnder interferometer (free-spectral range is 5.979 MHz). (\textbf{c}) Superposition of 10 cavity-ringdown signal scans for the fundamental TM mode with the corresponding decay time and the loaded and intrinsic Q factors. }
\label{fig:Fig2}
\end{figure*}

\begin{figure}[t!]
\centering
\includegraphics[width=\linewidth]{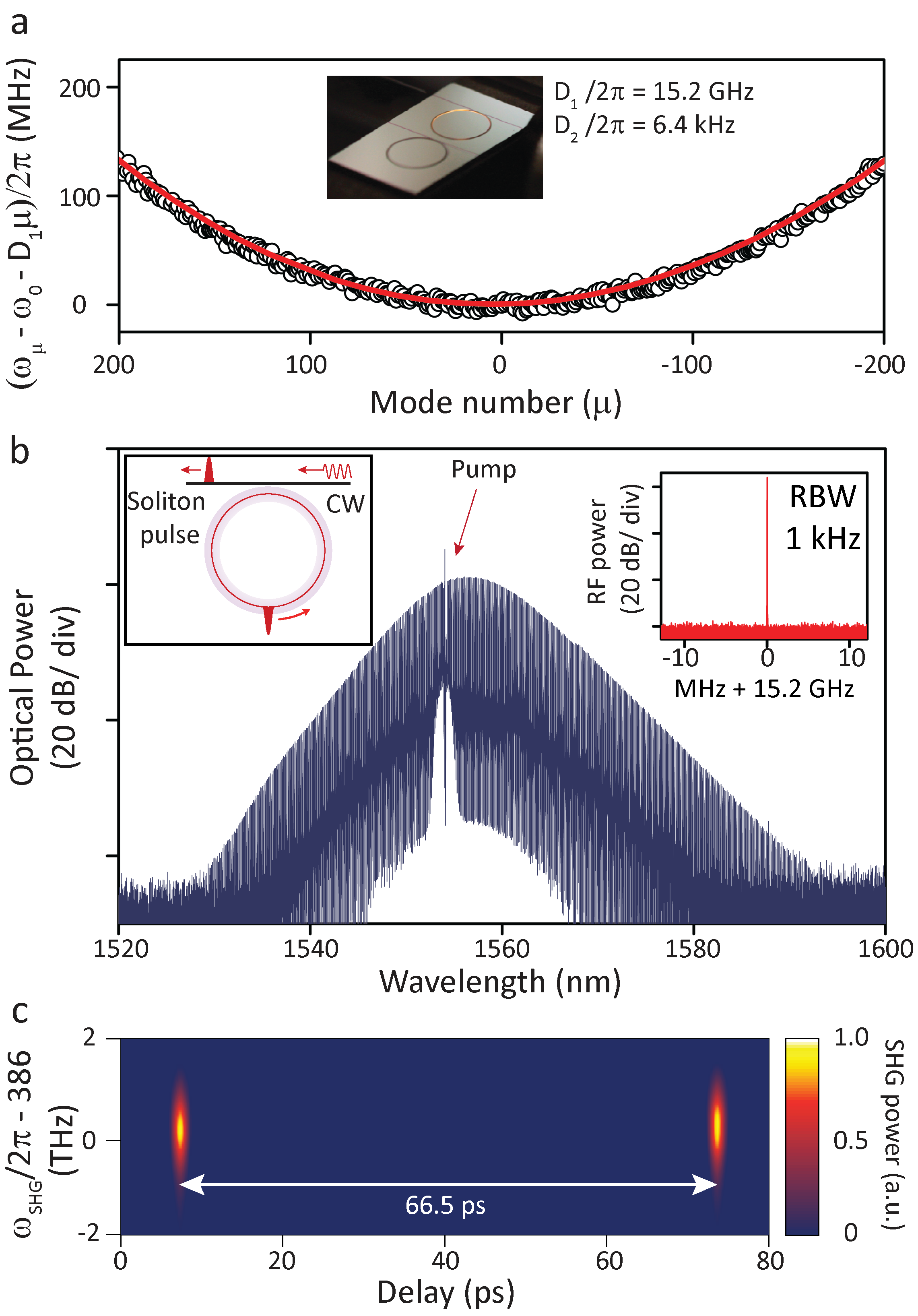}
\captionsetup{singlelinecheck=no, justification = RaggedRight}
\caption{{\bf{Demonstration of 15 GHz  repetition rate temporal solitons in an integrated optical microresonator}} (\textbf{a}) Measured mode frequencies plotted versus mode number for the fundamental TE mode family with linear dispersion removed. The red curve is a parabola showing that the dispersion is primarily second order over 400 modes. Legend gives the free-spectral range ($D_1$) for mode $\mu = 0$ and the second-order dispersion ($D_2$). The inset is a photograph of two ridge resonators. (\textbf{b}) Optical spectrum of soliton with pump line indicated. Left inset illustrates continuous-wave pumping to produce soliton mode locking of resonator. Right inset shows the electrical spectrum of the detected soliton pulse stream (RBW: resolution bandwidth) (\textbf{c}) FROG scan of single soliton state in 3b showing the single pulse signal with pulse period of 66.5 ps corresponding to resonator round-trip time.}
\label{fig:Fig3}
\end{figure}

\begin{figure}[t!]
\centering
\includegraphics[width=\linewidth]{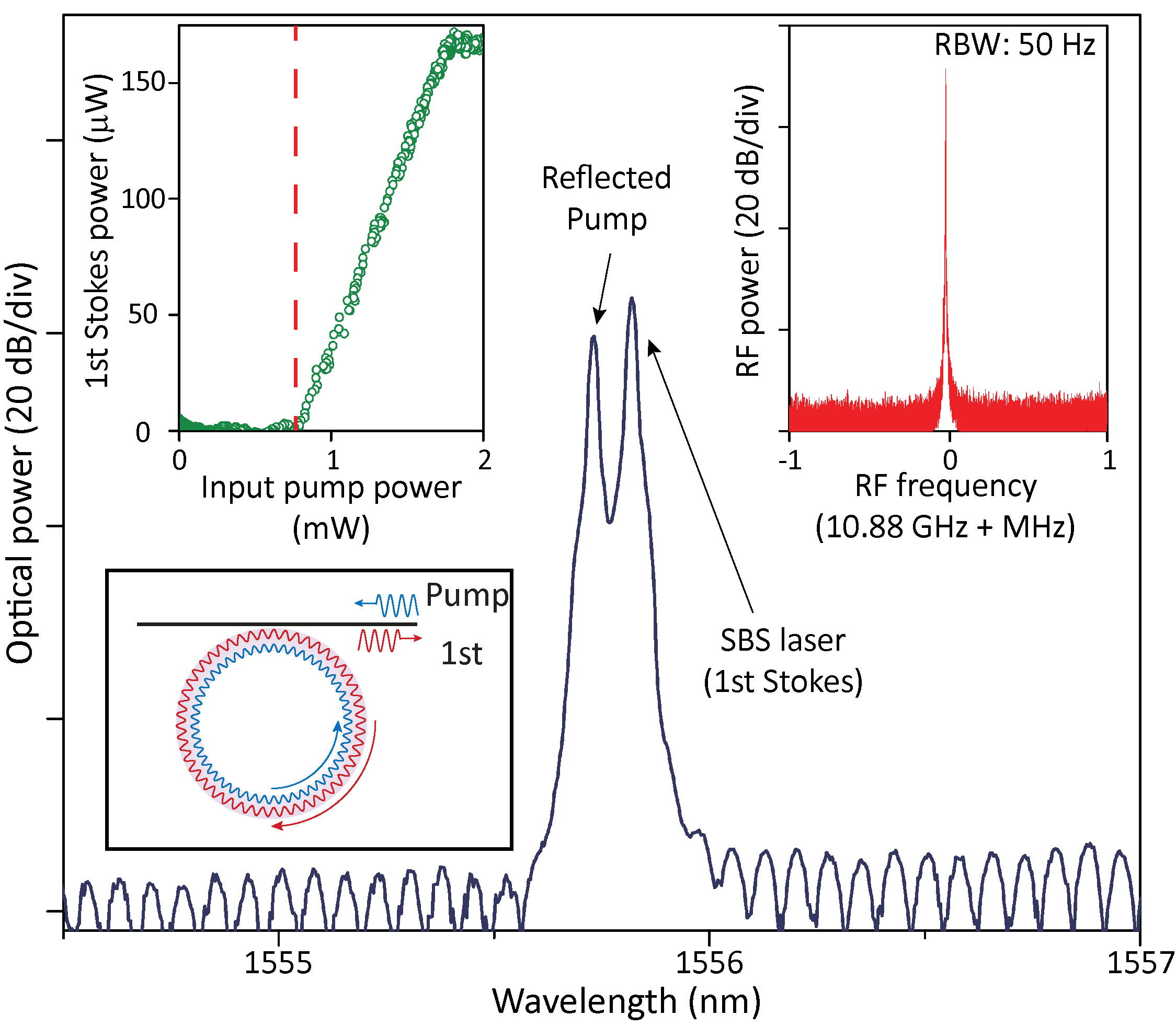}
\captionsetup{singlelinecheck=no, justification = RaggedRight}
\caption{{\bf{Demonstration of Brillouin lasing in an integrated optical microresonator}} Optical spectrum of Brillouin laser with pump and Stokes waves spectral peaks indicated. The left inset shows the output laser power vs. input coupled power and shows laser threshold behavior. All powers are in the waveguide. The clamping of the Stokes power occurs as a result of second-Stokes laser oscillation (i.e., Stokes cascade\cite{li2012characterization}). The right inset shows the microwave beatnote signal of the back-reflected pump and lasing signal. RBW is resolution bandwidth. Lower left inset shows pump (blue) and Brillouin laser action on first Stokes wave (red).}
\label{fig:Fig4}
\end{figure}

A cross-sectional rendering of the ridge resonator structure is provided as the inset in Fig. 1a. A mode field is also included. The resonator is composed of thermal silica and has a maximum thickness of approximately 8.5 microns. In Fig. 1b, a scanning electron microscope (SEM) image shows two of the silica ridge resonators connected by a common silicon nitride waveguide (false color red in the image). White arrows indicate how light guided in the silicon nitride couples into the ridge resonator where it circulates and then recouples to the waveguide. The silicon nitride waveguide has a thickness of approximately 250 nm. It is initially 3 to 3.5 microns in width at the edge of the wafer and is tapered to about 900 nm near the resonator so as to phase match to the resonator optical mode. The lightwave evanescently couples between the silicon nitride waveguide and the microcavity through the silica layer which supports the waveguide. The waveguide was designed in both straight versions and pulley versions to vary coupling strength\cite{Bowers_2014}. It is noted that because the waveguide has a higher index than the silica resonator, phase matching to resonator occurs when the waveguide supports only a single transverse mode. This improves waveguide-resonator coupling ideality\cite{Pfeiffer:PRA:2017}. Details on the fabrication process and phase matching of the waveguide to the resonator are provided in the Methods Section. SEM zoom-in and cross-sectional views of portions of the resonator are shown in Fig. 1c and in Fig. 1d. 

Spectral measurements of the integrated microcavity were performed by end-fire coupling to a tunable external cavity laser and monitoring of the transmission through the waveguide-coupled device as the laser is scanned. Approximately 25 \% of the power could be coupled from the lensed optical fiber to the silicon nitride waveguide. No attempt to improve coupling was made, but in the future a tapered waveguide end can be introduced to improve the coupling efficiency\cite{Lipson_2003}. The devices measured had a free-spectral-range (FSR) of approximately 15 GHz (ridge ring diameter of approximately 4.3 mm) and Fig. 2a presents a spectral scan containing over three FSRs. Several transverse modes appear in the scan. These transverse mode families were identified by measuring dispersion curves for the resonator modes as described below and comparing to numerical modeling. The waveguide used in this scan (and in Fig. 2b and 2c) was a pulley version. High-resolution scans of the fundamental TE and TM resonator modes are presented in Fig. 2b. The frequency scales in the scans are calibrated using a fiber Mach-Zehnder interferometer which appears as the green sinusoid in the lower half of each panel. Spectral calibration of the interferometer is performed using microwave sidebands generated by phase modulation of the laser. A linewidth fitting algorithm gives an intrinsic Q factor of 120 million for the TE mode and 205 million for the TM mode. To further confirm the high optical Q factor of the TM mode, cavity ring down was performed\cite{Lecaplain2016}. Fig 2c is a superposition of ten ring-down traces of the highest Q-factor measured from the fundamental TM mode family, and the ring-down gives an intrinsic Q factor of 230 million. The ring-down of the fundamental TM mode in Fig 2b gives an intrinsic Q factor of 216 million in close agreement with the linewidth data.

Microcombs and in particular soliton mode-locked microcombs \cite{Herr2014} represent a major new application area of microcavities. These tiny systems are being studied as a way to transfer large-scale frequency comb technology to an integrated photonic chip. They provide highly reproducible optical spectra, have achieved 2/3 and full octave span coverage\cite{Brasch2014,Pfeiffer2017} and have highly stable repetition rates\cite{Xu2015}. However, mode-locked pulse repetition rates that are both detectable and readily processed by electronics are required in all frequency comb systems in order to self-reference the comb\cite{Jones2000}. To achieve self-referenced octave-span operation in microcombs at practical power levels, frequency comb formation is divided into a THz-rate comb (micrometer-scale resonator diameter) and an electronics-rate comb (centimeter-scale resonator diameter)\cite{Spencer2017}. However, while the smaller-diameter, THz-repetition-rate, soliton combs have been demonstrated with integrated waveguides \cite{Li2016,Pfeiffer2017}, the large-diameter electronics-rate soliton microcomb has so far only been possible using discrete silica and crystalline resonators\cite{Herr2014,Xu2015}. Part of the challenge here is achieving a sufficiently high Q-factor to overcome the increased pumping volume of the larger electronics-rate soliton comb. Also compounding this problem are resonator design requirements imposed by the soliton physics. These include minimization of avoided mode crossings and anomalous modal dispersion. Attaining the combined features of ultra-high-Q factor to overcome a large optical pumping volume, while incorporating these resonator design features within an integrated microcavity has not been possible. Here, the integrated ridge resonator is used to demonstrate a new capability for an integrated design, 15 GHz soliton generation. The pumping power level is also low.

%Importantly, these devices are both miniature and have the performance required for self-referenced frequency comb systems. However, while integrated, electronic-rate microcombs have been demonstrated, these devices have so far operated only in the non-soliton regime. Non soliton microcombs tend to generate noisy and chaotic combs that are not suitable for practical, self-referenced microcomb systems. Accordingly, self-referenced microcomb systems have to date used discrete microcombs to generate the electronic-rate combs\cite{Spencer2017}.  

Using modeling and measurement techniques described elsewhere\cite{Herr2014,Xu2015}, the ridge resonator was designed to minimize avoided mode crossings as required for soliton generation. Along these lines, the fundamental TE mode family dispersion was characterized by measuring the frequency of all modes between 1530 nm and 1580 nm using an external cavity laser calibrated by the Mach-Zehnder interferometer described above\cite{Xu2015}. This data was then plotted by removing both an offset frequency ($\omega_0$) and a linear dispersion term proportional to the FSR ($D_1$) of the mode family at $\omega_0$ ($D_1/ 2 \pi =$ 15.2 GHz was measured). Modes are indexed using the label $\mu$ with $\mu = 0$ assigned to the mode at frequency $\omega_0$, which is also the mode that is pumped to generate the solitons. The resulting data are plotted in Fig. 3a. The red curve is a parabola showing that the mode family is dominated by second-order dispersion over the range of modes measured. A fitting gives $D_2/ 2 \pi =$ 6.4 kHz where $\omega_{\mu} = \omega_0 + D_1 \mu + D_2 \mu^2 / 2$. Significantly, there is no observable mode-crossing-induced distortion of the mode family, therefore making the mode family well suited for soliton formation\cite{Herr2014,Xu2015}. 

Solitons were triggered and stabilized using the power kick\cite{Brasch2014} and capture-lock technique\cite{Xu2016}. An optical spectrum is shown for a single soliton state in Fig. 3b. Visible in the spectrum is the optical pump wave. Despite the large (4.3 mm diameter) resonator required to attain the 15 GHz repetition rate, the ultra-high-Q ensured a low parametric oscillation threshold\cite{Kippenberg2004} power near 2 mW (waveguide coupled) and pump power as low 61  mW for stable soliton operation. To confirm the soliton repetition rate, the soliton pulse stream was detected and analyzed on an electrical spectrum analyzer. The electrical spectrum is provided as an inset to Fig. 3b and shows a measured repetition rate near 15.2 GHz. The pulse-like nature of the soliton stream was measured using the frequency resolved optical gate (FROG) method. The FROG data is presented in Fig. 3c. 

%Ultra-high-Q offsets the associated impact of increased mode volume on device operating power. In gyro systems 

%Optical gyroscopes present another major challenge for inegrated microresonators. Here, gyroscope performance is boosted by the combination of increased resonator size (increased Sagnac effect) and high-Q resonances (increased sensitivity). Indeed, discrete centimeter-scale silica and crystalline devices have recently achieved rotation rate sensitivities near the Earth's rotation rate using active\cite{Li:Optica:17} (Brillouin-laser) and passive\cite{Liang:Optica:2017} resonator approaches. In the active approach sub-Hertz linewidths of the Brillouin lasers result from the inverse quadratic scaling of their fundamental phase noise on Q factor\cite{li2012characterization}. This same phase noise scaling consideration is centrally important for microwave synthesizers\cite{Li2013} and microwave references\cite{Li2014} based upon high-coherence Brillouin lasers.

The Brillouin process has attracted considerable interest in micro devices \cite{eggleton2013inducing}. Brillouin laser action has been demonstrated in discrete resonators based on silica \cite{Tomes2009,Hansuek2012,li2012characterization} and CaF$_{2}$ \cite{grudinin2009brillouin}. Laser action has also been realized in integrated resonators using silicon \cite{Rakich2017Laser} and chalcogenide waveguides \cite{Eggleton:2017}.  However, reference sources\cite{Loh2015}, microwave synthesizers\cite{Li2013,Li2014} and Brillouin gyroscopes\cite{Li:Optica:17} require the highest possible optical Q factors for generation of narrow-linewidth signals and have so far relied upon discrete devices. Moreover, ultra-high-Q also allows these devices to function at low power despite their increased size. For example, discrete centimeter-scale silica\cite{Li:Optica:17} devices have recently achieved rotation rate sensitivities near the Earth's rotation rate using an active Brillouin-laser \cite{Li:Optica:17}. In this approach sub-Hertz Brillouin laser linewidths result from the inverse quadratic scaling of their fundamental phase noise on Q factor\cite{li2012characterization}. These narrow linewidths endow the device with high sensitivity. However, the gyroscope performance is also boosted by increased resonator size (increased Sagnac effect), and here the ultra-high-Q resonance provides an additional benefit by enabling Brillouin gyro operation at milliWatt-scale power levels. 

As a second device demonstration, the integrated ridge resonator is applied to generate high coherence Brillouin laser action. Devices were fabricated to phase match the Brillouin process when pumped near 1550 nm. Device diameters of approximately 6.0 mm were fabricated and tested. Fig. 4 shows the optical spectrum of the lasing Stokes wave. The weaker pump signal peak in the spectrum results from the need to collect the lasing Stokes wave in the propagation direction opposite to the pumping direction. Its strength is determined by residual backscattering in the measurement. The upper left inset in Fig. 4 shows the Stokes power versus pumping and gives a threshold power of approximately 800 $\mu$W in the waveguide. The upper right inset is the microwave beat signal between the pump wave and the Stokes wave. It has a high coherence as evidenced by the resolution bandwidth (RBW) and has a Hertz-level, short-term fundamental linewidth. 

Beyond the specific device demonstrations provided here, performance considerations relating to noise and fluctuations frequently require combination of high-Q and large mode volume that present challenges for scalable fabrication processes. For example, thermorefractive and photothermal  fluctuations\cite{Gorodetsky2004,Matsko:07} destabilize cavity frequency and vary inversely with resonator volume. Likewise, the fundamental timing jitter noise of a microcomb soliton stream is predicted to scale as $n_2 / V$ where $n_2$ is the Kerr nonlinear coefficient and $V$ is the mode volume\cite{Matsko2013}. To the extent these sources of fluctuation can be managed or reduced in active devices by increased cavity volume (or reduced $n_2$ as in the case of soliton timing jitter), higher Q factors can counteract the accompanying effect on pumping power. Finally, dual-comb spectrometers require reduced free-spectral-range (large diameter) devices so as to prevent under-sampling of chemical spectra. These larger microcomb devices also benefit from the highest possible Q factor so as to enable operation at practical power levels. 

In summary, we have demonstrated an integrated resonator having a record optical Q factor. Low-pump-power soliton generation at 15 GHz as well as high-coherence Brillouin laser action were demonstrated to illustrate functionality previously possible in only discrete devices, but nonetheless required for self-referenced microcomb photonic systems and for integrated systems requiring high-coherence signal sources. The waveguide material, PECVD silicon nitride, is among the most widely used waveguide materials in the photonics industry and provides a nearly universal interface to other photonic devices fabricated on
a common silicon wafer. 
\\

%\textbf{
%\begin{equation}
%\begin{aligned}\label{eqn:threshold}
%P_{th}=\frac{\pi n \omega_{0} A_{eff}}{4\eta n_{2}}\cdot\frac{1}{D_{1}Q^{2}}
%\end{aligned}
%\end{equation}
%Here $A_{eff} \sim 45 \mu m^2$ is the effective mode area, n is the refractive index, $n_{2}$ is the Kerr coefficient, $\eta = Q/Q_{ext}$ is the coupling efficiency where $Q_{ext}$ is the coupling Q factor, $\omega_{0}$ is optical frequency, and $D_{1}$ is the FSR in rad/s units (Lipson: cross-section$\sim 0.73 \times 2.50 \mu m$; Qi: $\sim 0.60 \times 3.00 \mu m$).   \\
%}

\noindent{\bf Methods}
\noindent Referring to Fig. 5, the fabrication process begins by growing a thermal silica layer on a high-purity float-zone silicon wafer. This layer is then patterned and wet etched using buffered hydrofluoric acid (HF) according to a process used to create wedge resonators\cite{Hansuek2012}. The resulting etched oxide forms a circular disk structure that defines the exterior of the resonator. A subsequent thermal oxidation grows the oxide layer beneath the oxide disk. This growth is followed by deposition of 500 nm of PECVD silicon nitride. Lithography and shallow ICP RIE dry-etch steps are then applied to partially etch the silicon nitride. Phosphoric acid is used to fully define the silicon nitride waveguide. A thin layer (20 nm) of silica is applied by atomic layer deposition to protect the silicon nitride waveguide during the final dry etch step below. A lithography/wet-etch step is then used to open an interior, ring aperture to the silicon substrate.  In the final step, xenon difluoride (XeF$_{2}$) etches the silicon through the interior ring aperture so as to create an optical cavity. 

\begin{figure}[h]
\centering
\includegraphics[width=\linewidth]{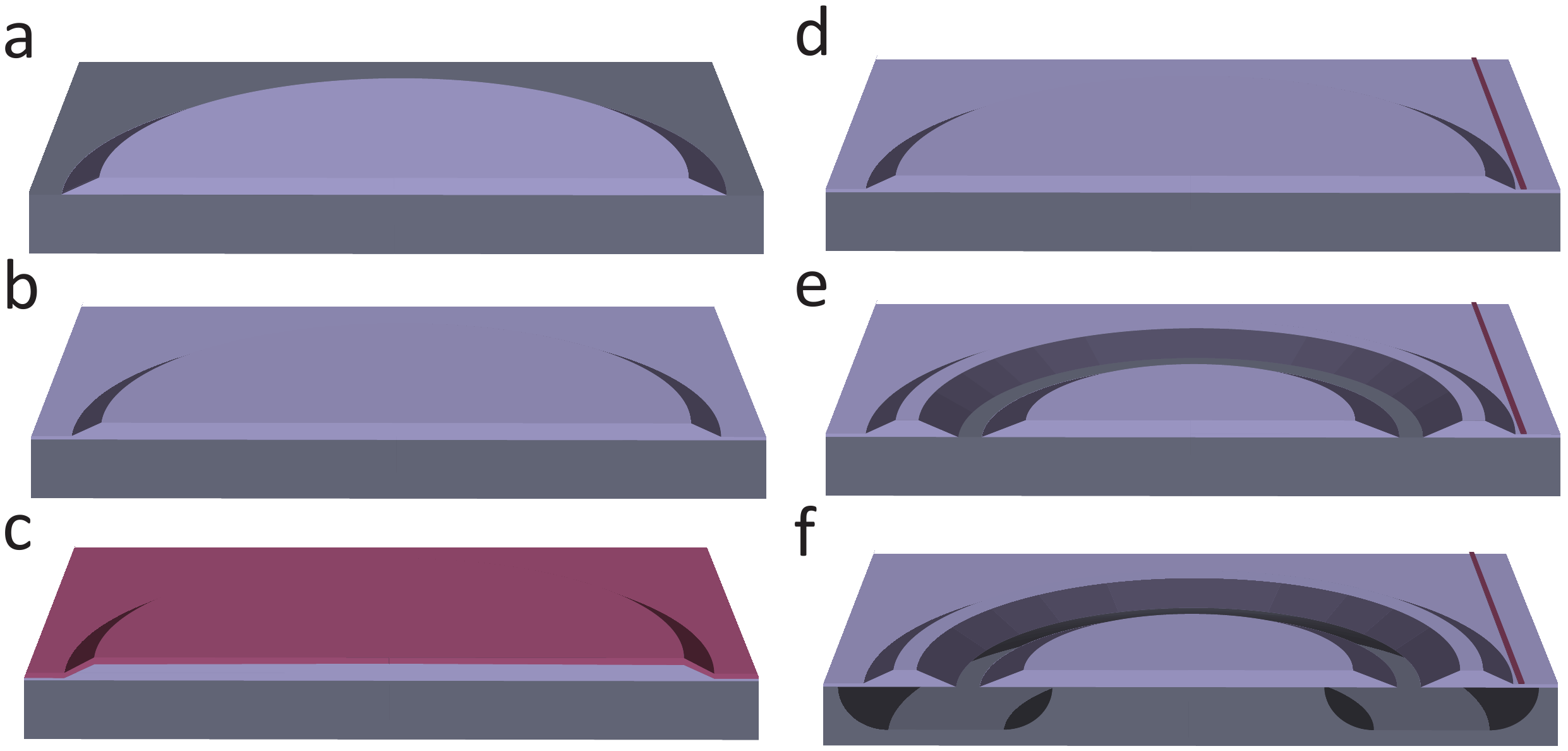}
\captionsetup{singlelinecheck=no, justification = RaggedRight}
\caption{{\bf{Fabrication process for integrated ultra-high-Q microresonator}} (\textbf{a}) Silica disk is defined on silicon by photolithography and HF etching of an initial silica layer. (\textbf{b}) Thermal oxidation grows a second oxide layer beneath the etched silica disk. (\textbf{c}) PECVD silicon nitride is deposited. (\textbf{d}) Silicon nitride waveguide is defined on silica layer by lithography and etching. (\textbf{e}) Lithography and wet etch of the silica defines a ring aperture to the silicon substrate. (\textbf{f}) XeF$_{2}$ etches the silicon through the ring aperture (also see inset in Fig. 1a). }
\label{fig:Fig_method}
\end{figure}

\begin{figure}[b!]
\centering
\includegraphics[width=\linewidth]{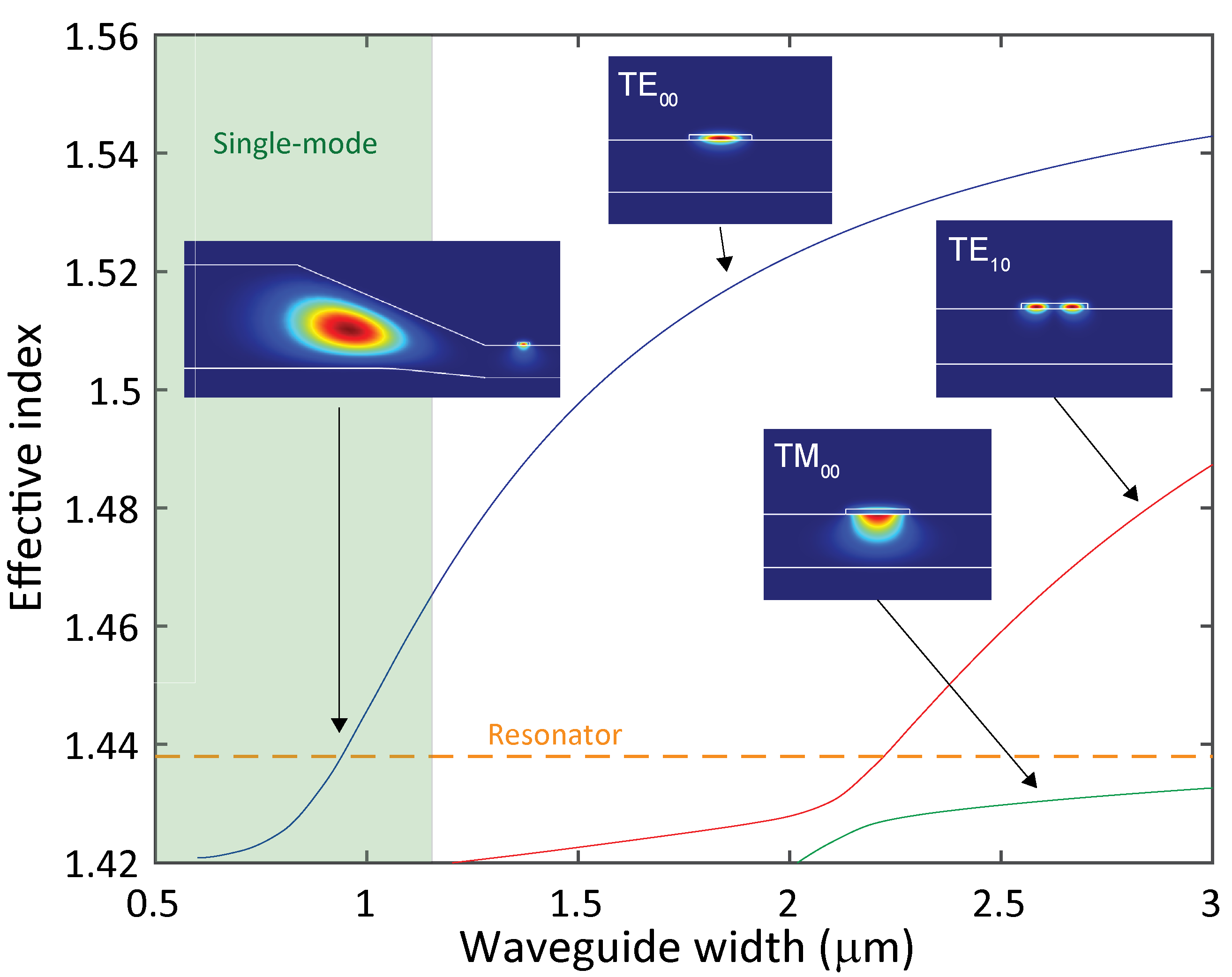}
\captionsetup{singlelinecheck=no, justification = RaggedRight}
\caption{{\bf{Design of a single-mode silicon nitride waveguide phase-matched to the silica resonator}} Calculated effective index of the silicon nitride waveguide modes versus the waveguide width. The waveguide thickness is 250 nm. The mode profiles for three spatial modes are provided as insets to the figure. Waveguides with width less than 1200 nm (green region) support only the fundamental TE mode (TE$_{00}$ mode). The phase-matching of the TE$_{00}$ mode to the fundamental resonator modes occurs near the waveguide width 900 nm.}
\label{fig:Fig5}
\end{figure}

In Fig. 6 the phase matching of the silicon nitride waveguide to the silica ridge resonator is studied by plotting the waveguide effective index versus the waveguide width (250 nm waveguide thickness). Three waveguide modes are studied. Below a width of approximately 1200 nm (green region), the waveguide supports only the fundamental TE mode (TE$_{00}$ mode). The required effective index to phase match to the resonator is indicated by a horizontal dashed line, and phase matching to the TE$_{00}$ mode occurs near a width of 900 nm. Insets are also provided to the illustrate the mode profiles both within the waveguide alone and the waveguide-resonator coupling section. An interesting and important feature of this system is that phase matching to the fundamental TE resonator mode occurs when the waveguide is single mode. As has been noted elsewhere\cite{Pfeiffer:PRA:2017} high-Q resonators tend to be overmoded and therefore phase matching to a waveguide can generally involve careful design to avoid non-ideality arising from the required larger waveguide dimensions. In the present case, the use of a waveguide material (silicon nitride) with index larger than that of the resonator material (silica) allows phase matching to occur when the waveguide is single mode.

\medskip

\noindent\textbf{Acknowledgment}
\noindent The authors thank Oskar Painter and Barry Baker for assistance with the PECVD silicon nitride process, Harry Atwater for assistance with silica atomic layer deposition, Matthew Hunt for assistance with electron beam microscopy, Yu-Hung Lai for technical assistance, and Andrey Matsko for helpful discussions on soliton timing-jitter noise. The authors also gratefully acknowledge the Defense Advanced Research Projects Agency under the DODOS program (Award No. HR0011-15-C-0055, Sub Award KK1540) and the Kavli Nanoscience Institute.\\

\noindent\textbf{Author contributions}
KYY, DYO, SHL and KV conceived the fabrication process and resonator design. KYY, DYO and SHL fabricated and tested the resonator structures. KYY, DYO and SHL conducted soliton and Brillouin laser measurement with assistance from QFY and XY. All authors analyzed the data and contributed to writing the manuscript. \\

\bibliography{Reference}
\end{document}